\documentstyle[12pt,aaspp4]{article}
     \raggedright

\received{2003 July 28}
     
\begin{document}

     \title{INTEGRAL Field Spectroscopy of the Extended Ionized Gas in Arp 220\altaffilmark{1}}
     %\vspace{0.5in}

     \author{Luis Colina\altaffilmark{2},
      Santiago Arribas\altaffilmark{3,4}, \&
     David Clements\altaffilmark{5}}

     %\vspace{0.25in}

     %\scriptsize
 \altaffiltext{1}{Based on observations with WHT operated on the\
      island of La Palma by the ING in the Spanish Observatorio del Roque de
      los Muchachos of the Instituto de Astrof\'{\i}sica de Canarias.}
     \altaffiltext{2}{Instituto de Estructura de la Materia (CSIC), Serrano 121,
                 28006 Madrid, Spain (colina@isis.iem.csic.es)}
      \altaffiltext{3}{Space Telescope Institute, 3700 San Martin Drive, Baltimore, USA 
         (arribas@stsci.edu). Affiliated with the Astrophysics Division, Space 
         Science Department of ESA.}
      \altaffiltext{4}{On leave from the Instituto de Astrof\'\i sica de Canarias -- 
         Consejo Superior de Investigaciones Cient\'{\i}ficas (CSIC)}  
      \altaffiltext{5}{Physics Dept., Imperial College, Prince Consort Road, 
                 London, SW7 2BW, UK; d.clements@ic.ac.uk}

     %\vspace{1in}

%      Based also on observations with the NASA/ESA Hubble Space Telescope,
%      obtained at the Space Telescope Science Institute, which is operated
%      by the Association of Universities for Research in Astronomy, Inc.
%      under NASA contract No. NAS5-26555.

     %\normalsize

     \begin{abstract}

Integral field optical spectroscopy with the INTEGRAL system has been
used to investigate for the first time the two-dimensional kinematic and 
ionization properties of the extended, warm ionized gas in Arp 220 over an area of 
75\farcs0 $\times$ 40\farcs0 (i.e. 28 $\times$ 15 kpc). The structure of the
ionized gas 
is divided into well identified regions associated with the X-ray emitting 
plumes and extended lobes, previously studied in detail by McDowell and collaborators. 
The overall ionization state of the warm gas in the plumes and lobes as traced 
by the [NII]/H$\alpha$ line is consistent with high velocity 
shocks expanding in a neutral ambient medium. Changes in the ionization state of the gas
along the major axis of the plumes are detected, in particular in 
the outer regions of the northwestern plume where the transition between the main stellar 
body of the galaxy and a broad, low surface brightness tidal tail is located. If the plumes 
are produced by a starburst-driven galactic wind,
the efficiency in the conversion of mechanical to radiation energy is a 
factor of at least 10 smaller than in galactic winds 
developed in edge-on spirals
with well defined rotation and outflowing axis.
The kinematic properties of the lobes with an average velocity of +8 km s$^{-1}$ (E lobe)
and $-$79 km s$^{-1}$ (W lobe) are to a first order in agreement with the predictions
of the merger scenario according to which the lobes are tidal-induced gas condensations 
produced during the merging process. The largest velocity gradients of 
50 km s$^{-1}$ kpc$^{-1}$ and velocity deviations 
of up to +280 km s$^{-1}$ and $-$320 km s$^{-1}$ from the systemic velocity are not 
associated with the plumes, but with the outer stellar envelope and broad tidal tails at
distances of up to 7.5 kpc, indicating that the large scale kinematics of the 
extended ionized gas in Arp 220, is most likely dominated by the tidal-induced motions, 
and not by galactic winds associated with nuclear starbursts.

     \end{abstract}


\section {INTRODUCTION}

Arp 220 is the prototypical cool (i.e. f$_{25}$/f$_{60}$ $<$ 0.2)
Ultraluminous Infrared Galaxy (ULIRG) with an infrared
luminosity L$_{IR}(8-1000 \mu$m) = 1.45
$\times 10^{12}$ L$_{\odot}$.  Arp 220 is also the nearest ULIRG with 
a redshift of 0.01813, and is thus an ideal laboratory to investigate with
 the physical processes taken place in ULIRGs.

Arp 220 has clear signs of being at the final stage of a merging
process. Optical images reveal extended, broad tidal tails (Arp 1966; Hibbard, Vacca, 
\& Yun 2000;
Surace, Sanders, \& Evans 2000), and high-resolution near infrared and
radio imaging show the presence of two nuclei separated by about 0.4 kpc (i.e. 0$\farcs$98
Carico et al. 1990; Graham et al. 1990; Baan \& Haschick 1995; 
Scoville et al., 1998, 2000, and
references therein). Moreover, Arp 220 is also extraordinarily rich in gas with
about 5 $\times$ 10$^9$ M$_{\sun}$ of molecular gas (M$_{H_2}$)
concentrated in the nuclear region of 250 pc in radius
(Scoville, Yun \& Bryant 1997).

The stellar and the cold gas components have been studied in detail
(Scoville, Yun, \& Bryant, 1997; Sakamoto et al., 1999; Hibbard,
Vacca, \& Yun, 2000; Surace, Sanders, \& Evans, 2000 and references
therein). The two nir/molecular nuclei and their associated disks --
which rotate in opposite directions -- are embedded in a kpc scale
molecular disk, and seem to rotate in the same sense as the 100 kpc scale HI
disk, suggesting a prograde-retrograde
encounter of two gas-rich spiral galaxies (Scoville et al. 1997).

The ionized gas component in the nuclear regions of Arp220 has been
studied on the basis of long slit observations along two position
angles (Heckman et al. 1990), and more recently
using integral field spectroscopy with INTEGRAL (Arribas, Colina 
\& Clements 2001, ACC01
hereafter). The detailed two-dimensional study by ACC01 has shown the
complex kinematical behaviour of the ionized gas within 2 kpc from the
dust-enshrouded nucleus, where three different components associated
with rotation and radial flows respectively, have been identified. 

The nuclear and extended X-ray emitting gas has also been investigated
with ROSAT PSPC images (Heckman et al. 1996), and more
recently with {\it CHANDRA} high spatial resolution X-ray imaging (Clements et al.
2002; McDowell et al. 2003, McD03 hereafter). The {\it CHANDRA} images show the presence of
four distinct, spatially separated emitting regions, the hard X-ray nucleus, the circumnuclear 
region, the extranuclear plumes, and the extended lobes, that could
be associated with different phenomema operating at different  physical scales , i.e. 
the dust-enshrouded AGN (nucleus), the (circum-)nuclear starburst and associated winds 
(circumnuclear regions and plumes, respectively), and the tidal forces produced during the 
interaction (extended lobes).   

However, a detailed kinematical and ionization study of the gas flows present in the 
extended ionized gas of Arp 220, and of how they are affected by the interaction process and
by the massive nuclear starburst, is still lacking and requires an observational 
technique able to obtain two-dimensional velocity fields and ionization maps of
the complex ionized gas structures detected in this galaxy over scales
of about 60$^{''}$ across.  Integral field spectroscopy (IFS, e.g., Arribas 
\& Mediavilla, 2000) is an ideal technique for such a study since it allows a simultaneous 
and complete mapping of both the kinematic and ionization properties of the warm
ionized gas over large areas (see Arribas \& Colina 2003 and references therein for previous
studies of ULIRGs using integral field spectroscopy).  

This paper presents the first two-dimensional spectroscopic study of the
extended ionized nebula in Arp 220 covering a region of 75$^{''} \times 40^{''}$. 
In the following sections, the stellar morphology ($\S$3.1), two-dimensional structure
($\S$3.2), ionization ($\S$3.3), and kinematics ($\S$3.4) of the ionized gas are presented. 
The origin of the extended ionized gas as produced by starburst-driven galactic winds 
($\S$ 4.1), and tidal forces during the merger process ($\S$ 4.2) are briefly 
discussed.  Throughout the paper, a distance of 
77.6 Mpc. At the assumed distance, 1$^{''}$ corresponds to 375 pc.

     \section{OBSERVATIONS AND DATA REDUCTION}

     Integral field spectroscopy of Arp 220 was obtained with the
     INTEGRAL system (Arribas et al. 1998) and the WYFFOS spectrograph
     (Bingham et al. 1994) in the 4.2m William Herschel Telescope.
     We used the wide field observing mode (i.e. bundle of fibers
     SB3), which consists of 115 fibers, each of 2.7$^{''}$ in diameter, and
     covering a 33.6$^{''}$ $\times$ 29.4$^{''}$ field-of-view.
     In addition, 20 fibers distributed in a circle of 45$^{''}$ in
     radius were used for simultaneous sky measurements. The
     spectra were taken using a 600 line/mm grating with an effective
     resolution of about 9.8 \AA\  and covering the 4500$-$7500 \AA\
     spectral range. The observations consisted of a mosaic of three
     pointings (nuclear, east and west nebula) covering almost completely 
     a rectangular area of 75$^{''} \times 40^{''}$ (see Figure 1).   
     The total integration time in each of the pointings,
     split into separated integrations of 1500 sec each, was 6000 sec. for 
     the nucleus, 15000 sec. for the east nebula, and 18000 sec. for the west nebula.
     The seeing (1\farcs1 - 1\farcs2) was very stable over the entire observing 
     period (two nights) and conditions were spectrophotometric.
         
     Data reduction followed the standard procedures applied to spectra
     obtained with two-dimensional fiber spectrographs that have 
     been explained elsewhere (Arribas et al. 1997, and references therein; see 
     also ACC01). The 2D distribution of the reduced spectra in the range of the
     redshifted H$\alpha$ and [NII] lines are presented 
     in Figure 2. The changes in the profiles of the lines and in the
     ionization conditions (i.e. [NII]/H$\alpha$ ratio) are obvious
     even from this figure alone.

     The absolute calibration was obtained observing the standard star SP1550+330 
     with the same system configuration and data reduction procedures as
     used for Arp 220. The relative calibration of the different pointings
     agrees to within 20\%, as obtained by comparing the regions of 
     overlapping. Taking into account additional uncertainties like the
     reliability of the H$\alpha$ and [NII] lines fits, the uncertainty in the
     absolute calibration is about 30\%.

     \section{RESULTS}

     \subsection {Stellar Structure}

The reduced two-dimensional spectra can be used to generate continuum and emission line
images by selecting specific wavelength windows. A red, narrow band (7425-7455 \AA) continuum
image (Figure 3) is generated for a direct comparison with the archival HST/WFPC2 
I-band (F814W filter; P.I. K. Borne) image of Arp 220 (Figure 1). 
Although with a poorer spatial resolution, the INTEGRAL 
red continuum image shows an overall stellar morphology in the nuclear
regions similar to the structure detected in the WFPC2 I-band image. 
In particular the nuclear dust lane that cuts the stellar distribution in half
is clearly visible, with the apparent optical nucleus, identified as the continuum
peak at these wavelengths, located northwest of it. The position of this
peak, that corresponds to $\alpha$(1950)= 15h 32m 46.8s and $\delta$(1950)= 23d 41' 09\farcs3
with an uncertainty of 1\farcs0 is offset from the infrared and radio nucleus by
1\farcs5$-$2\farcs0 (see ACC01).

An additional feature worth mentioning is the disrupted stellar light
distribution in the outer regions at distances between 4 and 11 kpc, i.e. 10\farcs0 
to 30\farcs0, from the nucleus (see Figure 3). These regions correspond most likely to
tidal tails and external regions of the stellar disks in the progenitor galaxies 
involved in the merger.
The faint northwestern tail
detected in the WFPC2 image (Figure 1) is clearly identified in the low resolution 
INTEGRAL continuum image (Figure 3) as a broad structure centered at about 20\farcs0
(i.e. 7.5 kpc) northwest of the nucleus and elongated almost parallel to the dust lane 
orientation 
for about 7 kpc. On the eastern side, the stellar envelope extends towards the SE up to a 
distance of about 30\farcs0 (i.e. 11 kpc). While the major axis of the nuclear stellar 
light distribution is given by the central dust lane along position angle PA 45 ($\pm$5), the 
extended stellar envelope, as seen also in the WFPC2 image and in previous deep ground-based 
images (Hibbard, Vacca \& Yun 2000), has a major axis oriented along PA 110, away 
from that of the 
dust lane and almost perpendicular to it. All these characteristics point to a stellar 
system in Arp 220 that as a consequence of the merging process, has not yet reached its
dynamically relaxed configuration.

\subsection{Extended Ionized Gas: Structure}

The major axis of the diffuse, optical ionized gas in Arp 220 has an end-to-end size of 
24 kpc, similar 
in length to the extended soft X-ray emitting regions (Heckman et al. 1996; McD03). The 
ionized gas distribution traced by the H$\alpha$ (Figure 4a) and [NII] 6584\AA\ 
(Figure 4b) emission lines clearly shows substructures and different components at various 
linear scales.
As originally noted by Heckman et al. (1996), and recently confirmed with higher spatial
resolution {\it CHANDRA} images (McD03), the structure of the extended X-ray emission
bears a strong morphological resemblance to the warm ionized gas traced by the
optical emission lines. For consistency
with the different components of the X-ray emitting gas already identified, we will also refer
in the rest of the paper to the circumnuclear regions, plumes and lobes, to identify 
the optical ionized gas regions at various scales. A summary of several observed and derived
properties for the different regions is given in Table 1.

\subsubsection{The Nucleus and Circumnuclear Region}

The nucleus and its surrounding circumnuclear region up to a radius of 2 kpc have already been 
investigated in more detail using INTEGRAL with a higher spatial and spectral resolution 
configuration (see ACC01 for details). The H$\alpha$ emission peaks at 0\farcs8 (i.e. 0.3 kpc) 
southwest of the 
optical stellar continuum peak, and coincides with the position of the
soft X-ray emission peak, which is displaced 1\farcs5 northwest of the hard X-ray emission source 
(Clements et al. 2002). The hard X-ray source agrees in position with the mid-infrared
 and brightest
radio source (Soifer et al. 2000), and it is consequently considered as the true nucleus of 
Arp 220. Therefore, the warm H$\alpha$ ionized gas associated with the nuclear hard X-ray 
source is enshrouded by the nuclear dust lane and appears distributed
with an apparent emission peak displaced 0.6 kpc northwest of the true ionizing nucleus.

\subsubsection{The Plumes}

On a scale of a few kiloparsecs, and up to distances of 4 kpc from the nucleus, two 
elongated high surface brightness regions, one on each side of the nucleus, 
with a total length of 7 kpc are identified. 
These regions, oriented along a position angle of 120 to 140 degrees coincide with the 
X-ray emitting NW and SE plumes (McD03). As for the soft X-ray emission,
the surface brightness of the outer part of the SE plume, not affected by the
absorption due to the central dust lane, is a factor four that 
of the NW plume (Figures 4a, 4b) indicating a clear intrinsic asymmetry in the properties 
of the warm and hot 
ionized gas in opposite directions of the dust-enshrouded nucleus. 
The close similary of the [NII]+H$\alpha$ and the X-ray 
plumes in relative surface brightness (i.e. the SE plume is brighter than the NW plume
by the same factor in both H$\alpha$ and X-ray emission), and morphology (see Figure 9 in
McD03) indicates that the warm and hot ionized gas components are closely related,
and most likely have the same physical origin (see $\S$4 for further discussion).

As mentioned above, the line emitting gas in the SE plume has on average a surface brightness 
four times that of the NW plume. There are several mechanisms that can be invoked to 
explain this asymmetry but none of them is clearly supported by the present set of data.
Assuming the same geometry for both sides of the plume and that they are
viewing the same radiation field coming from the nucleus, this enhancement in brightness could 
be explained by a 
density effect. Since H$\alpha$ and [NII] emission are a function of N$_e^2$, a difference 
of four in surface brightness would imply a difference of two in density, and therefore the 
gas in the SE plume would be a factor two more dense than the corresponding gas in the NW plume. 
However, there is no clear evidence for this systematic difference in the electron density of the
ionized gas. The measured [SII] 6717\AA~ to 6731\AA~ emission line ratio gives values of the
order of 1.0 $-$ 1.1, consistent with an average density of a few hundred particles per unit 
of volume in both SE and NW plumes. An alternative explanation could be that the NW plume is
seen through the dust lane, the intrinsic emission more absorbed, and consequently the 
surface brightness diminished relative to that of the SE plume. Although the non detection of the
H$\beta$ line in emission precludes any quantitative estimate of the local extinction using the Balmer decrement,
this scenario seems also not to be valid since there is no evidence for a radial
dependency of the NW to SE plume's relative surface brightness. In addition a factor four in
absorption would imply an almost constant extinction equivalent to 3 mag. in the visual over
distances of several kpc, which seems very unlikely in view of the non uniform structure
of the dust lane which also runs almost perpendicular to the orientation of the plumes.
A likely explanation for the difference in surface brightness is that the radiation 
field seen by the gas in the plumes, i.e. at distances of few to several kpc, is not 
symmetric with respect to the nucleus. This could be the case if 
the three dimensional structure of the dense dust and gas in the nuclear
regions surrounding the dust-enshrouded nucleus were such that would block the radiation
preferentially along a given direction.

At a lower surface brightness, there is a secondary plume (or filament) located south of 
the NW plume, and
elongated mostly parallel to it from ($-$5\farcs0, $-$4\farcs0) to ($-$12\farcs0, $+$8\farcs0),
with a total length of about 10\farcs0 (i.e. 3.75 kpc). The southwestern part of this secondary 
plume, brightest in [NII] light, emerges from the main circumnuclear emission peak 
(see Figure 4b) while the northwestern 
part of it, traced by the H$\alpha$ line, curves at ($-$10\farcs0, $+$1\farcs0) to reconnect 
with the NW plume at its tip (see Figure 4a).

\subsubsection{The Lobes} 

On scales of several kiloparsecs, at distances between 4 to 15 kpcs from the nucleus, 
the H$\alpha$ and [NII] light
distributions show the presence of two extended, limb-brightened lobes, one on 
each side of the nucleus. The west lobe (W lobe hereafter) is about a factor two brighter 
than the east lobe (E lobe hereafter), and
has an edge-brightened structure with a diameter of about 4.5 kpc centered at 7 kpc west of 
the nucleus. On the other hand, the low surface brightness E lobe centered at 9.4 kpc from 
the nucleus, is actually located southeast of the nucleus, and has an ellipsoidal 
edge-brightened shape with a major axis of about 9 kpc oriented along a position angle 
of 90 degrees (Figure 4b). 

An additional important feature is that the ionized gas in the lobes tend 
to avoid the regions occupied by the extended stellar envelope in the SE and by the broad
tidal tails detected in the NW (see Figures 3 and 4b). Moreover, the optical lobes also avoid
the neutral HI gas distribution that is located at about the same radial distances from the
nucleus but at a position angle of 50 degrees
(Hibbard, Vacca \& Yun 2000). On the contrary, as already known from previous studies
(Heckman et al. 1996; McD03), the overall structure of the optical lobes coincides with
the X-ray lobes. This spatial segregation between, on one side the warm and hot gas, and on
the other side the stellar and the cold neutral gas could favor the idea that the observed
stellar and gas distributions in the outer regions of a merger system like Arp 220 are 
the result of the tidal forces induced during the merging process. Therefore, the structure
and energetic of the ionized gas in these lobes are not associated with a starburst generated
wind but with the tidal forces and evolutionary state of the merger (see $\S$4.2 and
McD03 for further discussions).

\subsection{Extended Ionized Gas: The Ionization State}

\subsubsection{The Nucleus and Circumnuclear Region}

The ionization state of the circumnuclear region has been studied in more detail in ACC01.
However, the main results are briefly summarized here in order to get a complete
picture of the ionization of the gas at different scales. Due to the absorption caused by the
dust lane, the nucleus itself, identified as the hard X-ray, infrared and radio emission 
peak, is not 
detected as the brightest region in emission lines. As already mentioned, the peak of the 
H$\alpha$ and [NII] emission line distributions is located at
about 1\farcs5 NW (0.6 kpc) of the dust-enshrouded nucleus. The ionization structure of
the warm gas in the circumnuclear regions is very homogeneous with a few well-localized
regions where a change in the ionization state is detected (ACC01). On average, the overall 
ionization is consistent with LINER-like emission, or even Seyfert 2 around the peak 
ionized gas emission, if the lower limits to the [OIII]/H$\beta$ and [OI]/H$\alpha$ ratios are 
considered (see ACC01 for further details).

\subsubsection{The Plumes}

The ionization state of the plumes presents a complex structure with well-identified
regions along their major axis where the excitation conditions change on
the scale of kpc. The 
circumnuclear H$\alpha$ and [NII] main emission peaks located northwest of the 
dust-enshrouded nucleus are spatially coincident. However, 
the H$\alpha$ and [NII] secondary emission peaks are 
offset by about 1 to 1.4 kpc, and present a clear pattern. The [NII]
secondary peaks in the SE and NW plumes are located at distances of 
about 2 to 2.6 kpc from the main emission peak, respectively. On the other hand, the H$\alpha$
secondary peaks are located at the tips of the plumes, at distances of about 3.5 kpc. 
The average ionization state of the gas in the plumes is characterized by a 
[NII]/H$\alpha$ ratio of the order of 1.4 to 2.0, presenting a maximum value of 3 in the
region identified as the [NII] emission peak in the NW plume. Similarly, the 
[SII]/H$\alpha$ ratios have values of about 1, with a slightly larger value in the region
of the [NII] emission peak. Values this large for the [NII] and [SII] emission lines 
with respect to H$\alpha$ are consistent with high velocity shocks ($\sim$ 300 km s$^{-1}$)
expanding in a neutral, not preionized,
medium (Dopita \& Sutherland 1995).

The ionization state in the secondary plume/filament, south of the NW plume 
(see $\S$3.2.2 for details on its
orientation and length),
 also presents very interesting features. Ionization
gradients are clearly visible in this plume as traced by the regions of H$\alpha$ and 
[NII] high surface brightness. While the southwestern region of the secondary plume,
emerging from the main line emitting peak, emits mostly in [NII] light, 
the H$\alpha$ emission takes over in the curved northwestern region that reconnects 
this secondary plume with the tip
of the NW plume (see Figures 4a and 4b).  The change
in the ionization, in particular the enhancement of H$\alpha$ with respect to [NII]
appears in a region where an abrupt change in the velocity of the gas by about 
50$-$60 km s$^{-1}$ kpc$^{-1}$ is detected (see $\S$3.4.2 and Figures
5a and 5b).

\subsubsection{The Lobes} 

The ionized gas in the extended lobes is asymmetrically distributed with the W lobe
being on average a factor four brighter than the E lobe. The ionization state as traced
by the [NII]/H$\alpha$ ratio (1.1 to 1.4) is very uniform over the entire extension of 
the lobes, without any indications of ionization gradients or substructures at the present 
resolution (3\farcs, or about 1.1 kpc). As for the plumes, the detection of strong [NII] 
emission with respect to H$\alpha$ ([NII]/H$\alpha$ ratio between 1.1 and 1.4) is consistent 
with high velocity shocks evolving in a neutral, not pre-ionized, 
medium, although additional measurements of the [OIII]/H$\beta$ ratio would be needed to 
confirm this (see Dopita \& Sutherland 1995). The average density of the ionized gas in 
the W lobe is lower than in the plumes, and less than 100 electrons per cm$^{-3}$ as 
measured by the [SII] lines ratio. The weakness of the [SII] line emission in the SE lobe, 
precludes any measurement of the density in this region.

\subsection {Extended Ionized Gas: Velocity Field}

The large scale velocity field over the entire region of about 75\farcs0 $\times$
40\farcs0 has been traced by the [NII] 6584\AA\ emission line (see Figure 5).
The velocity field is rather complex, without any regular pattern but with substructures 
and large velocity gradients already visible on scales of 1$-$1.5 kpc that can be associated with 
different structures identified in the stellar and gas components. 
The systemic velocity (cz) of Arp 220 defined as the velocity of
the gas located in the region of the dust-enshrouded nucleus is 
5400 $\pm$ 20 km s$^{-1}$ (ACC01). The velocities mentioned in the following
sections will be given relative to this value.

\subsubsection{The Nucleus and Circumnuclear Region}

The velocity field in this region shows a very complex behaviour characterized
by the presence of at least three kinematically and spatially separated components
associated with rotation and radial flows. A detailed investigation of the gas motions
in these regions can be found in ACC01.

\subsubsection{The Plumes}

The velocity field of the SE and NW plumes show a very different structure.
The NW plume shows a smooth velocity field with an average approaching velocity
of $-$102 $\pm$ 37 km s$^{-1}$, and a very small velocity gradient of less
than 5 km s$^{-1}$ kpc$^{-1}$. On the contrary, the SE plume has an average
velocity of $-$4 $\pm$ 63 km s$^{-1}$, i.e. consistent with the systemic velocity, but
it presents large velocity gradients. In particular, while the inner parts of the SE plume, i.e.
closer to the dust-enshrouded nucleus, show a velocity similar to systemic, the outer regions
identified by the tip of the [NII] emission, and by the local H$\alpha$ emission peak, show 
velocity changes from about systemic, or even approaching, up to $+$100 $-$ 120 km s$^{-1}$, 
implying velocity gradients of about 70 to 100 km s$^{-1}$ kpc$^{-1}$. 

The velocity structure of the secondary plume/filament is such that the gas is 
moving with a velocity of $-$25 km s$^{-1}$ in the southwestern region connecting with the
[NII] and H$\alpha$ main emission peak, increases up to $+$85 km s$^{-1}$ at 
 10\farcs0 NW of the emission peak, and decreases again to about $-$50 km s$^{-1}$ 
in the region where this filament reconnects with the tip of the NW plume, where an 
approaching velocity of $-$115 km s$^{-1}$ has been measured.    

It this difficult to explain the very different kinematic behaviour observed in the 
plumes under the simple scenario of a symmetric biconical outflow having a common
origin in the dust-enshrouded nucleus. Differences in the physical conditions and/or 
kinematics of the interstellar medium, interactions between several gas systems moving 
with large relative velocities, and/or different relative orientations of the outflowing 
axis in the plumes have to be invoked before reaching any conclusion about the origin 
and evolution of the plumes (see discusion in $\S$4).

\subsubsection{The Lobes}

The velocity field of the ionized gas in the lobes indicates that the W lobe is approaching
with an average velocity of $-$79 $\pm$ 58 km s$^{-1}$ while the E lobe has a velocity
of +8 $\pm$ 79 km s$^{-1}$, consistent with the velocity of the system. Therefore, the
average velocity of the lobes is consistent, although with a larger dispersion, with 
that of their corresponding plumes,
independent of the overall physical scale of the two structures and of the different
orientations of their major axis.   
The W lobe has a velocity field with a maximum receding velocity of $+$30 km s$^{-1}$
and a minimum approaching velocity of about $-$160 km s$^{-1}$ at 25\farcs0
(i.e. 9.4 kpc) east of nucleus. The southern section of this lobe has
approaching velocities of up to $-$70 km s$^{-1}$. 
The ionized gas in the E lobe has also approaching 
velocities in the $-$25 km s$^{-1}$ to
$-$60 km s$^{-1}$ range with a maximum velocity of only $+$10 km s$^{-1}$, consistent
with the systemic velocity.  

Other important aspect of the ionized gas located in the limb-brightened lobes is its
different kinematical behaviour with respect to the corresponding gas associated, at least
in projection, with the southeastern stellar envelope and the broad, northwestern tidal 
tail. This velocity difference
amounts to $+$75 km s$^{-1}$ and $+$190 km s$^{-1}$ for the gas in the E and W lobes,
respectively (see $\S$3.4.4). These large velocity difference suggest that the
relatively high surface brightness lobes are physically separated and kinematically 
distinct from the low surface brightness, merger-induced stellar envelope and 
tidal tail (see $\S$4.2 for
a more detailed discussion).

%The ionized gas in the lobes is clearly redshifted with respect to the surrounding
%gas associated with the stellar envelope. In the W lobe, these velocity differences range
%from $+$210 up to $+$350 km s$^{-1}$, while in the E lobe the differences are much smaller 
%with values from $+$50 up to $+$120 km s$^{-1}$. Therefore the relatively high surface 
%brightness lobes are physically separated and kinematically distinct from the stellar 
%envelope and tidal tails. 

\subsubsection{Diffuse Gas Associated with the Stellar Envelope and Tidal Tail} 

In addition to the ionized gas associated with the X-ray emitting plumes and lobes, there are
low surface brightness ionized gas regions that seem to be 
associated with stellar structures that have been produced by the tidal forces generated
during the evolution of the merger. These regions, as
already mentioned, are identified as the extended stellar envelope southeast of the nucleus,
and as the broad, low surface brightness tidal tail northwest of the nucleus. 

Although the ionized gas in these regions represents a minor contribution to the overall
content, both in mass and energy, the largest deviations from systemic velocity are detected 
in these low surface brightness regions, and the velocity substructure of this diffuse gas 
follows precisely the changes detected in the 
structure of the stellar light distribution (see Figure 5c). The gas in the southwestern
 side of the central dust lane  moves with velocities 
of $-$ 110 km s$^{-1}$. Moreover, the gas associated with the broad stellar tidal tail 
located at about 20\farcs0 W of nucleus is moving with an approaching velocity of up to
$-$ 320 km s$^{-1}$,  representing the absolute minimum of the overall velocity field. 
On the other hand, low surface brightness gas associated with the southeastern stellar
envelope shows velocities ranging from $-$100 to $+$100 km s$^{-1}$ with a maximum
value of $+$280 km s$^{-1}$ at about 7\farcs0 NE and 17\farcs0 SE of the 
nucleus, representing the absolute maximum of the entire velocity field. 
In summary, the gas in these regions presents the largest velocity deviations 
 measured in the extended ionized regions of Arp 220 with a peak-to-peak value of 
600 km s$^{-1}$ at distances of about 5 to 7.5 kpc from the nucleus.

\section{DISCUSSION}

\subsection{Origin of the Plumes: Starburst-driven Galactic Wind Structures?}

Starburst-driven galactic winds, understood as generated by the 
collective effects of supernovae explosions and massive star winds generated in a 
(circum-)nuclear
starburst, are able to compress and sweep the ambient interstellar medium of
spiral galaxies with velocities of a few hundred km s$^{-1}$ along the direction of the 
maximum pressure gradient, i.e. in a direction (almost) 
orthogonal to the disk of the galaxy  (Heckman, 
Armus \& Miley 1990 and references therein). Evidence for such winds have been
reported in a number of edge-on spirals (Veilleux et al. 1994, Lehnert
\& Heckman 1996; Lehnert, Heckman \& Weaver 1999; Strickland et al. 2002) where the
finite thickness of the disk of the galaxy and the central location of the starburst
(or AGN) defines the direction of the maximum pressure gradient, 
and therefore of the wind. As already pointed out by McDowell and collaborators
(McD03), it is however 
very unlikely that the ordered interstellar medium and well defined geometry of spirals
would exist in advanced mergers such as Arp 220.
As already shown in the previous sections, the large scale gas/dust and stellar 
distribution do show in Arp 220 a complex
structure characterized by a central dust lane, a low surface brightness
stellar envelope and broad tidal tails at distances of several kpc. In addition,
the large scale two-dimensional velocity field does not show any regular pattern 
with a privilaged orientation or spin axis. All these characteristics support the idea 
that even if a galactic wind is generated in the
central region (i.e. inner kpc) as a consequence of a (circum-)nuclear starburst, 
changes in the velocity structure and physical conditions of the interstellar medium
in which the wind expands will disrupt it in an efficient way.
With all these caveats in mind, the consistency of the
starburst-driven wind scenario with the observed morphology and measured energetics
of the plumes is evaluated in the following paragraphs.

Previous high spatial resolution integral field spectroscopy of the nuclear 
region did detect the presence of three different 
kinematical components consistent with rotating gas in a circumnuclear disk plus 
a tilted, outflowing gas (ACC01). The SE and NW plumes are oriented 
roughly along the same direction as the outflowing circumnuclear gas, and with an
orientation similar to that of the spin axis of the circumnuclear rotating gas suggesting
that the plumes of ionized gas could be due to a galactic wind generated by the nuclear
starburst. However, the plumes detected in Arp 220 show properties very different from 
the wind-generated structures detected in edge-on spirals with central starbursts or AGNs
(see Table 2 for a summary of the different properties).  The morphology of the ionized gas 
outside 
the plane of the galaxy  in these nearby edge-on spirals indicates that the gas is 
concentrated in
filaments (NGC 3079), sheets/filaments (M82), hollow shells (NGC 253), or biconical
surfaces (NGC 2992). These structures indicate 
that the gas being ionized is mostly located in the surface of a bubble-like 
or cone-like structure, some with wide  opening angles ($\geq$ 90 degrees) as in
M82, NGC 2992 or NGC 3079, or some others with a more focussed beam ($\sim$ 25 degrees)
like in NGC 253. The 
Arp 220 plumes show on the contrary a centrally concentrated optical, and X-ray (McD03),
emission with, at the present resolution of
about 1 kpc, no evidence of being resolved. This puts an upper limit of about 15 to 20 
degrees for the opening angle if a cone-like or bubble-like structure as detected in
nearby spirals is assumed. Therefore, if the plumes in Arp 220 were a consequence of a 
starburst-driven wind, the outflow channel would have to be very narrow, even narrower than
the structure detected in NGC 253. Under the wind scenario, a narrow plume indicates a
well collimated outflow. In order to produce this structure, the wind would have to have 
a lateral expansion velocity much smaller than the outflowing velocity, and the axis of 
the maximum pressure gradient would have to be very stable over distances of at least 
few kpcs. Integral field spectroscopy with higher spatial resolution (0\farcs5) is needed 
in order to be able to resolve the plumes and investigate the velocity field across the
plume.  

Under the galactic wind scenario, the
shocks generated by the expanding wind produce an H$\alpha$ luminosity that is directly 
proportional to the preshock ambient density, and to the total surface of
the hollow cone or bubble, and has a strong dependence ($\propto$ V$_s^{2.4}$) on the 
velocity of the shock (see for instance Veilleux, Shopbell \& Miller 2001). Moreover, the total 
kinetic energy liberated by the starburst-driven wind is a few percent of the energy 
liberated by supernova explosions and massive stars winds, and therefore is directly 
proportional to the
star formation rate (SFR in M$_{\odot}$ yr$^{-1}$) as given by the expression 
E$_k$ = 9.4 $\times$ 10$^{54}$ $\times$ SFR ergs, 
if a continuous star formation over a period of 20 million years is
assumed (Colina, L\'{\i}pari \& Macchetto 1991). According to this, since Arp 220 is 
an ultraluminous infrared galaxy, about 50 times more luminous than the nearby spirals where
wind-related structures are detected,
the total mass and kinetic energy of the ionized gas in the plumes is expected to be
50 times that measured in spirals. However, this is not the case.  
The observed maximum velocity (not corrected for projection effects) is relatively low for
large kpc scale galactic winds, and the ionized gas mass is in the range measured in nearby 
galaxies (see Table 2). Moreover, if normalized by the infrared luminosity, the plumes in
Arp 220 (and even the plumes with the additional contribution from the lobes) are underluminous 
in X-ray and H$\alpha$ emission by factors of at least 10 with respect to galaxies
with prototype large scale, wind-induced structures including M82, NGC 2992 and NGC 3079 (see 
Table 2; see also Table 9 in Strickland et al. 2003). Only the small
(overall extend of 1.3 kpc) shell detected in NGC 253 has an H$\alpha$ to infrared
luminosity ratio similar to those measured in the plumes and lobes of Arp 220. 

In summary, the Arp 220's plumes as traced by the 
optical emission lines and X-ray emission, are
underluminous when compared with the galactic wind structures in nearby spirals. 
Therefore, if the starburst-driven wind scenario holds for Arp 220 plumes, 
the derived energetics indicates that the starburst-driven wind is much less efficient, 
by factors of at least 10 than in spirals with well developed wind-induced structures. 
Whether or not this behaviour is due to the  existence in ULIRGs of an ambient dense
interstellar medium with a complex three-dimensional structure and with a velocity field 
characterized by not having a privileged kinematical axis, requires further 
investigations and detailed modeling.

\subsection{Origin of the Lobes: Tidal-induced Structures?}

Although the starburst-driven wind interpretation could still hold as an explanation for the
origin of the ionized gas detected in the circumnuclear regions (ACC01) and  even in the
plumes, the large 
scale, i.e. several kpc size, two-dimensional kinematical structure 
presented here (see $\S$3.4) does not support this scenario as the origin of the 
edge-brightened lobes detected in X-rays and H$\alpha$. 
As already mentioned ($\S$3.4) the steepest 
gradients in the velocity field of the diffuse, extended gas are associated with substructures
identified in the stellar light distribution, in particular with the southwestern side of the
central dust lane, with the eastern extended stellar envelope, and with the broad, western 
tidal tail. Therefore, the three-dimensional gas and stellar structure of a merger galaxy 
such as Arp 220 makes 
it difficult to assume a stable axis for the direction of the maximum
pressure gradient that, as in edge-on spiral galaxies, would channel the gas from the central
regions into the outer envelopes of the galaxy at distances of 10 kpc, or larger.

In the following we explore the scenario of the lobes as a merger by-product
in Arp 220. As already pointed out by McDowell and collaborators (McD03), the different 
surface brightness and
orientation of the plumes and lobes suggest that these two regions are dynamically distinct
and points to a different origin. 
Within the framework of the merger scenario, the morphology 
of the X-ray/H$\alpha$ lobes, with their edge-brightened structure and orientation, is best 
fitted by the off-center,
face-on collision of two co-rotating disks with an impact velocity of less than a couple of
 hundred km s$^{-1}$ (McD03). 
According to the specific dynamical simulation of the Arp 220 merger (see McD03 for 
detailed explanations), the plumes and lobes are density condensations produced about 
150 million years after closest approach 
and 15 million years after the formation of the dense central region . Following this model 
further, the edge-brightened H$\alpha$
lobes are expanding outwards while also experiencing a general rotation, remnant of the 
original rotation of the two disks. The combination of the radial expanding flow and 
residual rotation of the disks is such
that the line-of-sight velocity along each of the lobe rims ranges from $+$185 to 
$+$25 km s$^{-1}$ 
(E lobe) and $-$175 to $-$20 km s$^{-1}$ (W lobe) in the frame of the galaxy (McD03). 
The data do not have enough
S/N to measure accurate velocities for each position along each of the rims, but are good 
enough to measure an average velocity for both E and W lobes.
The measured velocities corresponding to +8 $\pm$ 79 km s$^{-1}$ for the E lobe and 
$-$79 $\pm$ 58 km s$^{-1}$ for the W lobe, are to first order in agreement with the 
predictions but a more detailed comparison with the models is needed. Since the lobes
traced by the optical emission lines have a low surface brightness, low spatial resolution
integral field spectroscopy with 10m class telescopes is required in order to obtain
a significant improvement in the quality of the present spectra.

     \section{SUMMARY}

The complex two-dimensional structure, ionization state and velocity field of 
the warm ionized gas in the ultraluminous infrared galaxy Arp 220 has been investigated 
for the first time over a region of about 75\farcs0 
$\times$ 40\farcs0 (i.e. 28 $\times$ 15 kpc) using deep integral field optical spectroscopy 
with the INTEGRAL system. Arp 220 is the nearest ULIRG, and therefore the complexity
of its extended structure reaching the low
surface brightness levels of the ionized gas lobes previously detected in this unique target,
has been investigated with unprecedented detail. 
The main results from this study are:

(a) The structure of the ionized gas presents three distinct regions at different linear
    scales that show, as already known, an overall morphology similar to that of
    the X-ray emitting gas. The three different regions are: (a) the circumnuclear
    region up to a radius of about 5\farcs0 (2 kpc) already studied in ACC01, (b) the elongated 
    high surface brightness plume at radii between 5\farcs0 and 10\farcs0 (2 to 4 kpc), 
    and (c) the extended low surface brightness lobes at radii between 10\farcs0 and 35\farcs0
    (4 to 15 kpc).

(b) If the high surface brightness plumes are the result of a starburst-driven galactic
    wind, the measured energetics traced by the H$\alpha$ and X-ray luminosities indicates
    that it must be at least one order of magnitude less efficient than the bubble-like or
    cone-like galactic wind structures detected in nearby edge-on galaxies. Also, the
    opening angle of the plume in Arp 220 is very narrow ($\sim$ 20 degrees) when compared 
    with the wide opening angles typical of wind generated structures. These non-standard
    properties could be caused by  
    the complex three dimensional structure and kinematics of the gas and stellar
    components in advanced, luminous mergers such as Arp 220 that is very different 
    from the ordered galactic
    structure of edge-on spirals, where a rotation disk with a finite thickness defines the 
    axis of the outflowing wind, i.e. axis of maximum pressure gradient. This deserves proper
    investigations with detailed modeling and additional 2D spectroscopy.  

(c) The kinematics of the lobes with average velocities of +8 $\pm$ 79 km s$^{-1}$ and
    $-$79 $\pm$ 58 km s$^{-1}$ for the E and W lobe, respectively, are to first order in
    agreement
    with the predictions of the tidal-induced scenario proposed by McDowell and
    collaborators (McD03) by which the 
    gas condensations in the lobes are products of the merger that are animated by a 
    combination of
    continuing expansion and residual rotation from the original rotation of the two disks
    involved in the merger. This scenario has to be explored further by measuring the two
    dimensional velocity structure of the gas along the rims of the lobes with better 
    signal-to-noise spectra. Low spatial resolution integral field spectrographs 
    mounted on 10m class telescopes would be needed for this purpose. 

(d) The 2D velocity field shows its largest gradients (50 km s$^{-1}$ kpc$^{-1}$)
    and deviations (+280 km s$^{-1}$ to $-$ 320 km s$^{-1}$) from the systemic velocity
    in the low surface brightness ionized gas associated with the dust 
    lane, the southeastern stellar envelope and the northwestern broad stellar tidal 
    tail at distances of up to 7.5 kpc from the nucleus. Therefore, the velocity 
    field of the gas at distances of a few to several kpc from the nucleus and with 
    peak-to-peak deviations of 600 km s$^{-1}$, is associated with well defined 
    stellar structures induced by the merger, i.e. is dominated by tidal-induced 
    flows, and has no connection with a central starburst-driven galactic wind. 

Additional results are:

(1) The ionized gaseous plume is elongated along PA 120 (SE) to PA320 (NW) with the 
    southeastern side brighter than the northwestern side by a factor 4. An asymmetry 
    on the radiation field as seen by the gas in the plumes is the most likely explanation 
    for this difference. Alternative explanations like differences in the gas density and/or
    absorption effects through the plane of the galaxy are not supported by the present
    data. 

(2) The [NII] and H$\alpha$ emission peaks are separated by about
    1 kpc in both sides of the plume. The ionization structure is such that the
    [NII] emission is closer to the nucleus and the H$\alpha$ emission
    peaks at the tip of the plumes. The ionization state is consistent with it being
    produced by high velocity shocks expanding in a neutral ambient medium. 

(3) The change in the ionization state along the plumes, in particular the NW plume,
    traced by a change in the [NII]/H$\alpha$ ratio is interpreted as being produced
    by discontinuities in the velocity field and/or density of the gas at the interface
    of the main stellar body of the galaxy with the low surface brightness extended envelope. 
    
(4) The NW plume has a fainter, secondary component that runs almost parallel to the main 
    component and south of it. The inner part of this secondary plume is also dominated by the
    [NII] emission while the outer region emits in H$\alpha$. This change in the
    [NII]/H$\alpha$ ratio is interpreted, as before, as produced by variations in the 
    velocity field and/or density of the ambient gas.  

(5) The lobes show the same structure in both [NII] and H$\alpha$ with their peak
    emissions spatially coincident, within the present resolution. However, contrary to 
    the brightness distribution of the gas in the plumes, here the W lobe has a surface 
    brightness higher than the southeastern lobe by a factor 3, as previously detected in
    x-rays.

     \acknowledgements
     Luis Colina thanks the Instituto de Astrof\'{\i}sica de Canarias and the Space
     Telescope Science Institute for
     their hospitality and financial support. We also thank Luis Cuesta by his
     help using GRAFICOS.  Support for this work was provided by CICYT
     (Comisi\'on Interministerial de Ciencia y Tecnolog\'{\i}a) through
     grants PB98-0340-C01, PB98-0340-C02 and AYA2002-01055. Dave Clements acknowledges
     funding by PPARC. Email exchanges and discussions with Dr. Strickland are also acknowledged.

% and by NASA through grant number
%     GO-06346.01-95A from the Space Telescope Science Institute.

\clearpage

\begin{deluxetable}{ccccccccc}
\scriptsize
%\tablewidth{33pc}
\tablecaption{Properties of the Extended Ionized Regions in Arp 220}
\tablehead{
Region\tablenotemark{a} & Distance\tablenotemark{b} &
PA\tablenotemark{a}
& Extent 
& F$_{obs}$(H$\alpha$)\tablenotemark{c} 
& log L(H$\alpha$)\tablenotemark{d}
& M$_{H\alpha}$\tablenotemark{e} & V$_{average}$\tablenotemark{f} & E$_k$\tablenotemark{g} \\ 
 & (kpc) & & (kpc $\times$ kpc) & (10$^{-15}$ erg s$^{-1}$ cm$^{-2}$) & (erg s$^{-1}$) & 
(10$^4$ M$_{\odot}$) 
& (km s$^{-1}$) & (10$^{49}$ erg)\\ }
\startdata
Circumnuclear & 0 & -- & 1.1 $\times$ 1.1 & 31.0 & 40.4 & 84.6 & 5397 $\pm$ 74 & -- \nl
NW\_Plume\_H$\alpha$ & 3.5 & 320 & 1.5 $\times$ 1.9 & 1.9 & 39.1 & 4.4 & 5305 $\pm$ 39 & 4 \nl
NW\_Plume\_[NII] & 2.6 & 320 & 1.3 $\times$ 1.3 & 1.0 & 38.8 & 2.4 & 5291 $\pm$ 21 & 3 \nl
SE\_Plume\_H$\alpha$ & 3.5 & 125 & 2.4 $\times$ 2.4 & 5.4 & 39.6 & 14.6 & 5404 $\pm$ 58 & -- \nl
SE\_Plume\_[NII] & 2.0 & 125 & 2.3  $\times$ 2.3  & 4.5 & 39.5 & 12.3 & 5389 $\pm$ 61 & -- \nl
W\_Lobe & 7.0 & 260 & 7.1 $\times$ 7.5 & 9.3 & 39.8 & 21.5 & 5321 $\pm$ 54 & 14\nl
E\_Lobe & 10.1 & 114 & 8.3 $\times$ 5.6 & 2.1 & 39.2 & 5.4 & 5408 $\pm$ 76 & -- \nl
\tablenotetext{a}{Regions labeled as NW\_Plume\_H$\alpha$ and NW\_Plume\_[NII] are the regions
identified in the NW plume with the H$\alpha$ and [NII] emission peaks, respectively. The same 
applies for
regions SE\_Plume\_H$\alpha$ and SE\_Plume\_[NII]}
\tablenotetext{b}{Distances for the plumes are given as the distance between the nucleus and
the local H$\alpha$ and [NII] peak emission in the plumes. For the lobes is the distance between
the nucleus and the midpoint of the structure traced by the [NII] emission. The nucleus is 
identified here as the central bright source, even if this does not correspond with the true 
dust-enshrouded nucleus of the galaxy, not visible in the optical.}
\tablenotetext{c}{Flux in units of 10$^{-15}$ erg s$^{-1}$ cm$^{-2}$
A$^{-1}$. Uncertainties of
10\%$-$15\%.}
\tablenotetext{d}{Luminosity not corrected by internal extinction.}
\tablenotetext{e}{Mass of ionized gas assuming an average density of 100 cm$^{-3}$.}
\tablenotetext{f}{The systemic velocity is 5400 $\pm$ 20 km s$^{-1}$ from previous
 INTEGRAL measurements(ACC01). Velocities as measured and not corrected for projection 
 effects.}
\tablenotetext{g}{Kinetic energy assuming the mass and average velocity given in previous
 two columns and a systemic velocity of 5400 km s$^{-1}$.}

\enddata
\end{deluxetable}

\begin{deluxetable}{ccccccccccc}
\scriptsize
\tablewidth{44pc}
\tablecaption{Extended Ionized Gas: Arp 220 versus Galactic Winds in Spirals}
\tablehead{
Galaxy & Distance\tablenotemark{a} &
logL$_{IR}$\tablenotemark{b} & L$_{H\alpha}$/L$_{IR}$\tablenotemark{c} & 
L$_{X}$/L$_{IR}$\tablenotemark{d}
& Extent\tablenotemark{e} 
& M$_{H\alpha}$\tablenotemark{f} & V$_{max}$\tablenotemark{g} & Structure\tablenotemark{h} 
& $\theta$\tablenotemark{ii} & References\tablenotemark{j}\\ 
 & (Mpc) & (L$_{\odot}$) & (10$^{-5}$) & (10$^{-5}$) & (kpc) & (10$^4$ M$_{\odot}$) & 
(km s$^{-1}$) & & (Degrees) & \\ }
\startdata
Arp 220 Plumes & 77.6 & 12.16 & 0.14  & 0.6 & 7 & 34 & 110 & jet & 15-20 & 1, 2 \nl
Arp 220 Lobes & 77.6 & 12.16  & 0.16 & 0.3 & 24 & 27 & 160 & hollow shell & 70 & 1, 2 \nl
M82 & 3.0 & 10.52 & 16 & 2.5  & 2.4  & 66 & 300 &   filaments & 90 & 3, 4 \nl
NGC 253 & 3.0 & 10.46 & 0.37 & 1.5 & 1.3 & 1 & $<$100 & hollow shell & 20-26 & 3, 4, 5 \nl
NGC 2992 & 31 & 10.46 & 2740 & - & 5.6  & 10$^3$ & 200 & biconical & 125-135 & 6 \nl
NGC 3079 & 16 & 10.49 & 2.7 & 6.2 & 3 & 10 & 2000 & filaments & 90 & 3, 4, 7, 9\nl
 \tablenotetext{a}{Distances as given in reference (3) for M82, NGC 253 and NGC 3079, and
  in reference (6) for NGC 2992.}
\tablenotetext{b}{Luminosities as given in reference (3) for M82, NGC 253 and NGC 3079, and
  in reference (6) for NGC 2992.}
\tablenotetext{c}{H$\alpha$ luminosities corrected for internal reddening except for 
Arp 220 where the observed values given in Table 1 are used. } 
\tablenotetext{d}{L$_X$ indicates the soft X-ray luminosity of the extended, diffuse halo
 X-ray structure as given in reference (2) for Arp 220 and reference (4) for M82, NGC 253,
 and NGC 3079 and scaled to the distances assumed in column (2)}
\tablenotetext{e}{End-to-end size of the bubble or cone shaped structure, not corrected 
for projection effects.}
\tablenotetext{f}{Mass of the ionized gas assuming an average density of 100 cm$^{-3}$.}
\tablenotetext{g}{Line of sight velocities given (in absolute values) relative to the 
velocity of the system.}
\tablenotetext{h}{Structure defined from ground-based H$\alpha$ imaging and/or high-resolution
X-ray imaging.}
\tablenotetext{i}{Opening angle of the structures as given in the references, or derived from
the published images (M82)}
\tablenotetext{j}{(1) this work; (2) McDowell et al. (2003); (3) Heckman et al. (1990);
(4) Strickland et al. (2003);
(5) Strickland et al. (2000); (6) Veilleux, Shopbell \& Miller (2001); 
(7) Veilleux et al. (1994); (9) Cecil, Bland-Hawthorn \& Veilleux (2002)}

\enddata
\end{deluxetable}

\normalsize
\clearpage

\figcaption{Archival WFPC2 I-band image of Arp 220 showing the central regions as
      well as the extended, low-surface brightness envelope. The overlays indicate the
      area covered by the three independent pointings with INTEGRAL. 
      North is up and East is left.} 

     \figcaption{Spatial distribution of the spectra around the redshifted
      H$\alpha$+[NII] spectral region obtained after the data reduction process is
      completed, including internal cross-calibration of the three pointings using the 
      spectra on the overlapping regions.}

     \figcaption{INTEGRAL image of the stellar light
     distribution in the central regions and faint extended envelope of Arp 220 as traced
     by a line-free, narrow-red continuum window (7425-7455 \AA) generated using all 
     SB3 spectra. Although with a poorer resolution, note that the overall structure is
     similar to that detected in the higher resoslution HST image (Figure 1). The contours
     are in arbitrary units and represent the 5\%, 7\%, 12\%, 18\%, 26\%, 36\%, 45\%, 53\%,
     72\%, 92\% of the innermost contour. As in figure 1, North is up and East is left.}

     \figcaption{INTEGRAL images of the ionized gas distribution as traced by the 
      H$\alpha$ (a) and [NII] 6583\AA\ lines (b), respectively. The three different regions 
      identified
      in X-rays, bright circumnuclear region, elongated plumes, and extended, diffuse lobes, 
      are easily identified here. Changes in the
      ionization structure of the plumes, traced by the offsets between the H$\alpha$ and 
      [NII] emission peaks, are clearly visible. The contours for H$\alpha$ represent
      surface brightness levels of 0.1\%, 0.8\%, 1.9\%, 3.8\%, 6.9\%,
      12.1\%, 20.8\%, 35.4\%, 59.6\%, and 100\% of the maximum level (6.8 $\times$ 
      10$^{-16}$ erg cm$^{-2}$ s$^{-1}$ arcsec$^{-2}$). The orientation is the same as 
      in previous figures.}
 %{\bf SANTIAGO: los contornos escala logaritmica?
 %     no parece: el de 
 %     Ha tiene 10, 77.4, 189.9, 377.5, 690.4, 1212.4, 2083.1, 3535.6, 5958.4 y 10000} 

     \figcaption{Velocity field of the extended ionized gas obtained from the bright
      [NII] 6583\AA\ emission line. The different figures represent the correspondance
      of the velocity structures with the stellar, [NII], and H$\alpha$ light distributions,
      represented by contours in figures (a), (b) and (c), respectively.}

%%%UCP%%%
%\newpage
%\plotone{f1.eps}
%\newpage
%\plotone{f2.eps}
%\newpage
%\plotone{f3.eps}
%\newpage
%\plotone{f4a.eps}
%\newpage
%\plotone{f4b.eps}
%\newpage
%\plotone{f5a.eps}
%\newpage
%\plotone{f5b.eps}
%\newpage
%\plotone{f5c.eps}


\clearpage
     \begin{thebibliography}{}


     \bibitem[Arp 1961]{ARP61} Arp, H., 1966, ApJS 14, 1
     \bibitem[Arribas {\it{et al.}}~1998]{ARR98} Arribas, S., et al. 1998 SPIE, 3355, 821
     \bibitem[Arribas \& Colina 2002]{AC02} Arribas, S., \& Colina, L. 2002, ApJ 573, 576
     \bibitem[Arribas, Colina, \& Clements 2001]{ACC01} Arribas, S., Colina, L., \&
              Clements, D. 2001, ApJ 560, 160 (ACC01)
     \bibitem[Arribas \& Mediavilla 2000]{ARR00} Arribas, S., Mediavilla, E. 2000
              in  'Imaging the Universe in Three Dimensions: Astrophysics with Advanced 
              Multi-wavelength Imaging Devices', Ed. W. van Breugel and J. Bland-Hawthorn,  
              ASP, Conf. Ser., vol. 195, 295.
     \bibitem[Arribas {\it{et al.}}~1997]{ARR97} Arribas, S., Mediavilla, E., 
              Garc\'{\i}a-Lorenzo, B. \& del Burgo, C. 1997, ApJ 490, 227
     \bibitem[Baan \& Haschick 1995]{BH95} Baan, W.A., \& Haschick, A.D. 1995, ApJ 
              454, 745
     \bibitem[Bingham {\it{et al.}}~1998]{BI94} Bingham, R.G., Gellatly, D.W., Jenkins,
              C.R., \& Worswick, S.P. 1994 SPIE, 2198, 56
%     \bibitem[Borne, Bushouse, Colina \& Lucas 2001]{BBCL} Borne, K. D., Bushouse, H., 
%              Colina, L., \& Lucas, R. 2001, ApJ Suppl., submitted
%     \bibitem[Borne et al. 2000]{BBLC} Borne, K. D., Bushouse, H., Lucas, R. A., \&
%              Colina, L. 2000, ApJ 529, L77
     \bibitem[Carico et al. 1990]{CAR90} Carico, D.P., Graham, J.R., Matthews, K., 
              Wilson, T.D., Soifer, B.T., Neugebauer, G., \& Sanders, D.B. 1990, 
              ApJ 349, L39
     \bibitem[Cecil et al. 2001]{CE01} Cecil, G., Bland-Hawthorn, J., Veilleux, S., \&
              Filippenko, A. 2001, ApJ 555, 338
     \bibitem[Clements et al. 2002]{CL02} Clements, D.L., et al. 2002, ApJ 581, 974
%     \bibitem[Colbert et al. 1998]{COL98} Colbert, E.J.M., Baum, S.A., O'Dea, C.P., \&
%             Veilleux, S. 1998, ApJ 496, 786
     \bibitem[Colina, L\'{\i}pari \& Macchetto 1991]{CLM91} Colina, L.,  L\'{\i}pari, S.,
             \& Macchetto, F. 1991, ApJ 379, 113
%     \bibitem[Colina, Arribas \& Borne 1999]{CAB} Colina, L., Arribas, S., \&
%              Borne, K. D. 1999, ApJ 527, L13
%     \bibitem[Colina, Arribas, Borne \& Monreal 2000]{CABM} Colina, L., Arribas, S.,
%              Borne, K. D., \& Monreal 2000, ApJ 533, L9
     \bibitem[Graham et al. 1990]{GRA90} Graham, J.R., Carico, D.P.,  Matthews,
              K., Neugebauer, G., Soifer, B.T., \& Wilson, T.D., 1990, ApJ 349, L39
     \bibitem[Heckman, Armus \& Miley 1991]{HAM90} Heckman, T. M., Armus, L., \&
              Miley, G.K. 1990, ApJ Suppl. 74, 833
     \bibitem[Heckman et al. 1996]{HE96} Heckman, T.M., Dahlem, M., Eales, S.A., 
             Fabbiano, G., \& Weaver, K. 1996, ApJ 457, 616
     \bibitem[Hibbard, Vacca \& Yun 2000]{HVY00} Hibbard, J.E., Vacca, W.A., \&
              Yun, M.S. 2000, AJ 119, 1130
     \bibitem[Lehnert \& Heckman 1996]{LH96} Lehnert, M.D., \& Heckman, T.M. 1996, 
              ApJ 462, 651
     \bibitem[Lehnert, Heckman \& Weaver 1999]{LHW99} Lehnert, M.D., Heckman, T.M.,
              \& Weaver, K.A. 1999, ApJ 523, 575
     \bibitem[McDowell et al. 2003]{McD03} McDowell, J.C., et al. 2003, ApJ 591, 154 (McD03) 
     \bibitem[Sakamoto et al. 2000]{SAK00} Sakamoto, K., Scoville, N.Z., Yun, M.S.,
              Crosas, M., Genzel, R., \& Tacconi, L.J. 2000, ApJ 514, 68
     \bibitem[Scoville, Yun \& Bryant 1997]{SYB97} Scoville, N.Z., Yun, M.S., \&
             Bryant, P. 1997, ApJ 484, 702
     \bibitem[Scoville et al. 1998]{SCO98} Scoville, N.Z., Evans, A.S., Dinshaw, N.,
             Thompson, R., Rieke, M., Schneider, G., Low, F.J., Hines, D., Stobie,
             B., Becklin, E., \& Epps, H. 1998, ApJ 492, L107
     \bibitem[Scoville {\it{et al.}} 2000]{SCO00} Scoville, N.Z. et al. 2000,
              AJ, 119, 991
     \bibitem[Soifer {\it{et al.}} 2000]{SOI00} Soifer, B.T., Neugebauer, G., Matthews, K.,
             Egami, E., Becklin, E.E., Weinberger, A.J., Ressler, M., Werner, M.W.,
             Evans, A.S., Scoville, N.Z., Surace, J.A., \& Condon, J.J. 2000, ApJ 509, 523 
     \bibitem[Strickland et al. 2002]{ST02} Strickland, D.K. et al. 2002, ApJ 568, 689
     \bibitem[Strickland et al. 2003]{ST03} Strickland, D.K. et al. 2003, Astro-ph/0306592
     \bibitem[Surace, Sanders \& Evans 2000]{SSE00} Surace,J.A., Sanders, D.B., \& Evans,
              A.S. 2000, ApJ 529, 170 
     \bibitem[Veilleux et al. 1994]{VEI94} Veilleux, S. et al. 1994, ApJ 433, 48
     \bibitem[Veilleux, Shopbell \& Miller 2001]{VSM01} Veilleux, S., Shopbell, P.L.,
              \& Miller, S.T. 2001, ApJ 121, 198
     \end{thebibliography}
     \end{document}